# Effect of Chemical Composition on Enthalpy of Evaporation and Equilibrium Vapor Pressure

*Vladimir Kh. Dobruskin*[*]

Pashosh St. 11, Beer-Yacov 30700, Israel.

E-mail: dobruskn@netvision.net.il

**RECIVED DATE**

[*]To whom correspondence should be addressed. Tel: (972)-506-864 642; Fax: (972)-77-531 3448.

**Abstract**

Proceeding from the Clausius-Clapeyron equation, the relation is derived that establishes a correlation between the partial enthalpy of evaporation from binary solutions, concentrations of components, and equilibrium vapor pressures. The difference between enthalpies of evaporation of components from solutions and those from the pure liquids, $\Delta(\Delta H)$, depends on the chemical nature and concentrations, $X$, of solutions. The effect of concentrations on $\Delta(\Delta H)$ makes different appearances in ideal and non-ideal solutions, although, as a whole, $\Delta(\Delta H)$ increases with the growth of concentration of the second component. A model is introduced, which considers $\Delta(\Delta H)$ as the sum of energetic changes of three sequential stages: passage of molecules from the bulk liquid into the surface layer, exit of the molecules on the outer side of the interface, and the following desorption into the gas phase. In the framework of the model, the main contribution to enthalpy of evaporation comes from the processes in the surface layer. It is suggested that adsorption from solutions, which changes the chemical composition of the surface layer with respect to that of the bulk solution, determines, to great extent, the difference in the forms of the curves $\Delta(\Delta H)=f(X)$ for ideal and non-ideal solutions.

**Introduction**

The present paper deals with solutions of organic substances in equilibrium with the vapor phase. The theory of solutions is an integral part of the classical chemical thermodynamics;[1-8] the review of the literature may be found in monographs of Moelwyn-Hughes and Ben-Naim.[7,8] As a rule, the influence of mixture composition on equilibrium properties in the framework of classical thermodynamics is found from the assumed effect of concentrations on the chemical potentials and the relevant activity coefficients.[1-7] We believe that, except the classical approach, the concentration effects may be described on the basis of the Clapeyron equation.

The fundamental Clapeyron equation, $dp/dT = \Delta S/\Delta V$, is a thermodynamically exact equation that contains no approximations. Here, $p$ is the equilibrium vapor pressure, $T$ is the temperature of phase transition, $\Delta S = \Delta H/T$, $\Delta H$ and $\Delta V$ are the changes of entropy, enthalpy and volume, respectively, associated with a change of phase. In practical problems one may neglect the molar volume of the condensed phase relative to the molar volume of gaseous phase and approximates the latter by the ideal gas equation, $V = RT/p$, were $R$ is the gas constant. Then the Clapeyron equation results in the Clausius-Clapeyron approximation:

$$\frac{dp}{p} = \frac{\Delta H}{RT^2} dT \qquad (1)$$

Taking the latent heat as constant over a sufficiently small temperature interval, the equation integrates giving the integral Clausius-Clapeyron equation, which relates the temperature dependence of the vapor pressure to the change of enthalpy of the phase transition. The Clausius-Clapeyron equation has been checked experimentally over a wide range of conditions in experiments on the vapor pressure of solids and liquids and in measurements of melting curves. All the experiments have shown it to be obeyed to a high order of accuracy. Its validity provides one of the most direct tests of the truth of the second law. A solution of the Clapeyron equation where the above approximations are relaxed has been given by Shilo and Chez.[9]

Although it is a general practice to use the Clapeyron equation for the account for the temperature dependence of equilibrium parameters, it is quite surprising that its application for the description of concentration effects, to our knowledge, is nowhere available in the literature. We believe that the application of the Clapeyron equation as a basis for a study of the concentration-dependent properties provides the new insight into interactions in solutions, reveals the dependence of the enthalpy of evaporation (condensation) upon concentrations of solutes and gives the correlation between variations of enthalpy and shifts of equilibrium pressures.

**Main equation.**

Being an immediate corollary of the fundamental first and second laws, the Clapeyron equation is valid for any equilibrium systems and, in particular, for the following: (1)- for the system which contains a binary liquid mixture of volatile organic compounds in equilibrium with the saturation vapor and (2)- for the reference system which contains only the pure component. The latter may be thought of as a solution with zero concentration.

To distinguish between the systems and components, we shall apply the following notations: Subscripts "$m$" and "$ref$" refer to the mixture and the reference pure liquid, respectively, while subscripts "1" and "2" applied to any symbol indicate the first and second components of solutions. In the case of the first component, the Clausius-Clapeyron equation takes the following forms: (1) for the solution

$$\frac{dp}{p_{m1}} = \frac{\Delta H_{m1}}{R} \frac{dT}{T^2} \tag{2}$$

and (2) for the pure first component at the same temperature

$$\frac{dp}{p_{ref1}} = \frac{\Delta H_{ref1}}{R} \frac{dT}{T^2}$$ (3)

where $p_{m1}$ and $p_{ref1}$ are the equilibrium pressures of the first component over the solution and over the pure component, respectively, $\Delta H_{m1}$ is enthalpy of evaporation of the first component from the solution, and $\Delta H_{ref1}$ is that from the pure liquid. Note that $p_{ref1}$ is just the saturation pressure of the pure first liquid, $p_{s1}$. Subtracting eq 3 from eq 2 one obtains

$$d \ln \frac{p_{m1}}{p_{s1}} = \frac{\Delta(\Delta H)_1}{R} \frac{dT}{T^2}$$ (4)

Here $\Delta(\Delta H)_1 = \Delta H_{m1} - \Delta H_{ref1}$ is the difference between the enthalpies of evaporation of the first component (i) from the solution and (ii) from the pure liquid. The completely analogous relation is obtained for the second component. Further the subscripts "1", "2", and "$m$" will be omitted on the understanding that the equations will be valid for either component of solutions. To integrate eq. 4, the dependence of enthalpies on temperature must be introduced. Although either enthalpies of evaporation, $\Delta H_m$ or $\Delta H_{ref}$, varies with temperature, their difference $\Delta(\Delta H)$ is supposed to be practically independent of temperature, since the temperature changes, to great extent, cancel out each other by the operation of subtraction. In any case, the assumption about the constancy of difference, $\Delta(\Delta H)$, is closer to the observation than that in respect to either term, $\Delta H_m$ or $\Delta H_{ref}$, taken alone. Taking the difference of latent heats as the constant, an integration of eq 4 leads to $\ln p/p_s = -\Delta(\Delta H)/RT + C$, where C is the integration constant. The constant of integration here is equal to zero, because at any temperature $p$ approaches $p_s$ when $\Delta(\Delta H)$ tends to zero. Finally, one obtains

$$RT \ln \frac{p}{p_s} = -\Delta(\Delta H) \qquad (5)$$

As the equilibrium vapor pressure of solution is a function of its concentration, equation 5 enables one to study the effect of concentrations of solutes on the enthalpy of evaporation. It is the basic equation for the treatment experimental data, which shows how the latent heat of evaporation must be changed to produce the observed variations in the equilibrium pressures. The equation gives the straightforward results without the necessity of introduction of the additional (and sometimes vague) postulates about interactions in the system of interest; the nonideality of solutions is taken into account immediately by the computed values of the enthalpy changes making the application of the activity coefficients unnecessary.

**Experimental data**

Experimental data in the form of equilibrium partial vapor pressures versus mole fractions of solutes, $X$, are taken from three literature sources.[10-12] The equilibrium data for mixtures of (1) Carbon Tetrachloride and Benzene, (2) Carbon Tetrachloride and Ethyl Acetate, (3) Chloroform and Acetone, (4) Carbon Disulfide and Acetone, (5) Carbon Disulfide and Methylal, and (6) Ethanol and Water are given in *International Critical Tables*;[10] those for (7) Aniline and Cyclohexane, (8) Aniline and Toluene, (9) Ethyl Bromide and Ethyl Iodide, (10) Aniline and Methylaniline, and (11) Hexane and Heptane are presented by Kogan, Fridman, and Kafarov.[11] The mixture of (12) Chloroform and Ethanol are studied by Raymond.[12] The values of $\Delta(\Delta H)$ are computed by eq 5.

**Variations of enthalpy of evaporation of ideal solutions**

The changes in enthalpies of evaporation versus logarithm of mole fractions ln$X$ in the ideal solutions are presented in Figures 1-6. Note that the coordinate system adopted in the present study differs from that in the theory of solutions. Since for binary solutions there is an elementary relation between the mole fractions of the first ($X_1$) and second ($X_2$) components, $X_1=1-X_2$, properties of solutions are usually plotted against one of the coordinates, either $X_1$ or $X_2$. In contrast to the regular practice, the values of $\Delta(\Delta H)$ for each component are plotted here against its own mole fraction.

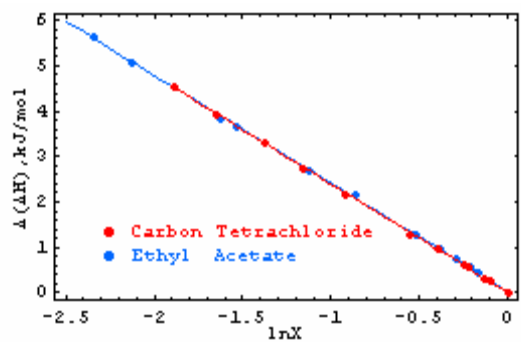

**Figure 1.** Variations in enthalpy of evaporation $\Delta(\Delta H)$ of components at 323.14 K in a mixture of Carbon Tetrachloride and Ethyl Acetate versus logarithm of mole fractions ln$X$.

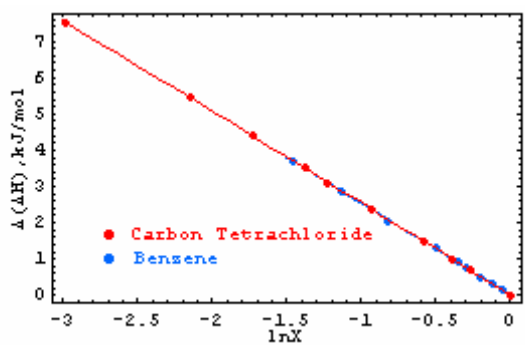

**Figure 2.** Changes in enthalpy of evaporation $\Delta(\Delta H)$ of components at 323.14 K in a mixture of Carbon Tetrachloride and Benzene versus logarithm of mole fractions ln$X$.

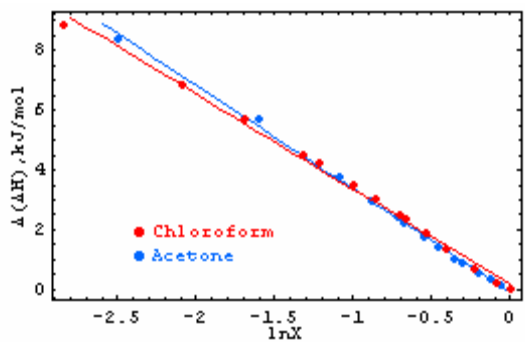

**Figure 3.** Variations in enthalpy of evaporation $\Delta(\Delta H)$ of components at 308.32 K in a mixture of Chloroform and Acetone versus logarithm of mole fractions ln$X$

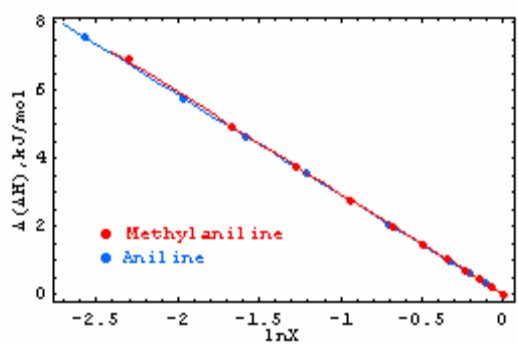

**Figure 4.** Variations in enthalpy of evaporation $\Delta(\Delta H)$ of components at 368.15 K in a mixture of Aniline and Methylaniline versus logarithm of mole fractions ln$X$.

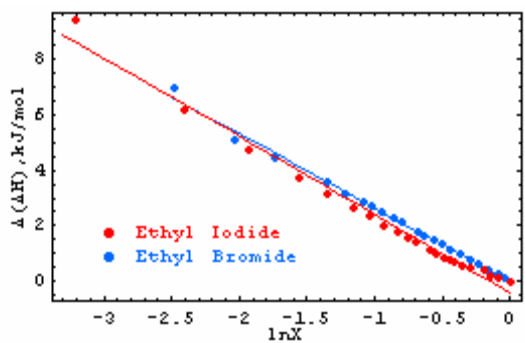

**Figure 5.** Variations in enthalpy of evaporation $\Delta(\Delta H)$ of components at 303.15 K in a mixture of Ethyl Bromide and Ethyl Iodide versus logarithm of mole fractions ln$X$.

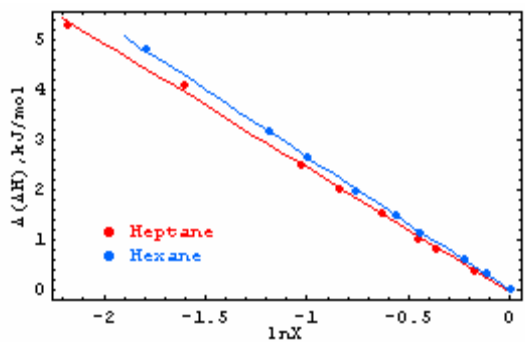

**Figure 6.** Variations in enthalpy of evaporation $\Delta(\Delta H)$ of components at 303.15 K in a mixture of Hexane and Heptane versus logarithm of mole fractions $\ln X$.

For example, the red line and points in Figure 1 describe $\Delta(\Delta H)$ of $CCl_4$ against the mole fractions of $CCl_4$, whereas $\Delta(\Delta H)$ of Ethyl Acetate (blue line and blue points) is given versus the mole fractions of Ethyl Acetate. Such a coordinate system is convenient when one compares values of $\Delta(\Delta H)$ for the solution components. One may see that (1) variations of enthalpies of each components increase with the growth of concentration of the second component and (2) the plots related to the first and second components practically coincide with one another for all mixtures presented in Figures 1-6.

Note that values of $\Delta(\Delta H)$ for the binary solutions, $\Delta(\Delta H_1)$ and $\Delta(\Delta H_2)$, relates to the different compounds: $\Delta(\Delta H_1)=\Delta H_1^s - \Delta H_1^0$ and $\Delta(\Delta H_2)=\Delta H_2^s - \Delta H_2^0$, where the superscript "0" indicates the pure component and superscript "s" denotes the solution. Nevertheless, $\Delta(\Delta H_1)=\Delta(\Delta H_2)$ at the equivalent mole fractions of components. It is not a trivial fact; further it will be shown that this equality is valid only for the ideal solutions. The difference in enthalpies of evaporation $\Delta(\Delta H)$ originates in the nature of solutions: a part of intermolecular bonds between molecules of the first or second components (1-1 or 2-2 bonds), existing in the pure substances, is broken and substituted by intermolecular bonds 1-2 between different molecules. Therefore, the number of such bonds determines values of $\Delta(\Delta H)$. For example, consider $\Delta(\Delta H_1)$ at $X_1=0.2$ and $\Delta(\Delta H_2)$ at $X_2=0.2$. Since $\Delta(\Delta H_1)=\Delta(\Delta H_2)$, the numbers of 1-2 bonds in these solutions are the same, although the molar ratios between the first and second components are completely different and equal to 0.2/0.8=1:4 and 0.8/0.2=4:1, respectively.

Variations of enthalpies of evaporation versus logarithm of mole fractions $\ln X$ for these systems are approximately described as follows:

$$\Delta\left(\Delta H\right) = a + b \ln X \tag{6}$$

where a (J mol$^{-1}$) and b (J mol$^{-1}$) are the coefficients of the straight line. Points in Figures 1-6 correspond to the vales of $\Delta(\Delta H)$ calculated from experimental observations. The coefficients $a$ and $b$ are found as a least-squares fit to the calculated values of $\Delta(\Delta H)$. The functions of $\Delta(\Delta H)=f(X)$ with the fitted coefficients are plotted in Figures 1-6 by colored lines. One may see that there is a good agreement between the points and the approximate equation. The substitution of eq 6 into eq 5 results in the relation for calculating the equilibrium pressures:

$$p = p_s Exp\{-\frac{a}{RT}\} \times X^{-\frac{b}{RT}} \tag{7}$$

The coefficients of equations 6 and 7 for the ideal solutions are given in Table 1.

Table 1. Parameters of the ideal solutions

| solution | component | T, K | coefficients | | | |
|---|---|---|---|---|---|---|
| | | | $a$, J mol$^{-1}$ | $b$, J mol$^{-1}$ | $e^{-a/RT}$ | $-b/RT$ |
| $CCl_4$-$CH_3COOC_2H_5$ | Ethyl acetate | 323.14 | 35.896 | -2374.6 | 0.987 | 0.884 |
| | $CCl_4$ | | 5.8103 | -2390.0 | 0.998 | 0.890 |
| $CCl_4$-$C_6H_6$ | Benzene | 323.14 | 17.319 | -2516.6 | 0.994 | 0.937 |
| | $CCl_4$ | | 20.989 | -2519.7 | 0.992 | 0.938 |

| CHCl$_3$ - OC(CH$_3$)$_2$ | Chloroform | 308.32 | 163.35 | -3183.8 | 0.938 | 1.242 |
|---|---|---|---|---|---|---|
| | Acetone | | -94.440 | -3458.0 | 1.037 | 1.305 |
| C$_2$H$_5$Br- C$_2$H$_5$I | Ethyl bromide | 303.15 | -37.771 | -2675.5 | 1.015 | 1.061 |
| | Ethyl iodide | | -449.10 | -2836.0 | 1.195 | 1.125 |
| C$_6$H$_5$NH$_2$ - C$_6$H$_4$CH$_3$ NH$_2$ | Aniline | 368.15 | 3.6618 | -2933.5 | 0.999 | 0.958 |
| | Methylaniline | | -11.707 | -2975.7 | 1.004 | 0.972 |
| C$_6$H$_{14}$ - C$_7$H$_{16}$ | Hexane | 303.15 | -24.4937 | -2933.49 | 1.010 | 1.064 |
| | Heptane | | -67.9824 | -2506.05 | 1.027 | 0.994 |

For mixtures of Hexane-Heptane, Aniline-Methylaniline, and Carbon Tetrachloride-Benzene both exp$\{-a/RT\}$ and $b/RT$ are close to unity; these mixtures are approximately subject to the Raoult law, $p=p_sX$ , and are referred to as the ideal solutions. In the cases of solutions of Ethyl Acetate-Carbon Tetrachloride, Ethyl Bromide-Ethyl Iodide, and Acetone-Chloroform, exp$\{-a/RT\}\neq1$; they are, nevertheless, subject to the Henry law, $p\approx KX$, at the low concentrations of the second component, where $K\neq p_s$ is the Henry constant. Such mixtures are usually called the ideal dilute solutions.

**Variation of enthalpies of evaporation of non-ideal solutions**

For the non-ideal solutions, the variation of $\Delta(\Delta H)$ in Figures 7-11 are presented as functions of mole fractions; in these cases, the logarithmic scale of concentrations does not improve the representation. One may see that the curves of $\Delta(\Delta H)=f(X)$ for ideal and non-ideal solutions differ in shape. For non-ideal solutions, the cross-points and flex-points appear on the curves, and values of $\Delta(\Delta H)$ of the individual components of solutions do not coincide with one another. Hence, there are different resistances to evaporation of each component of the solutions.

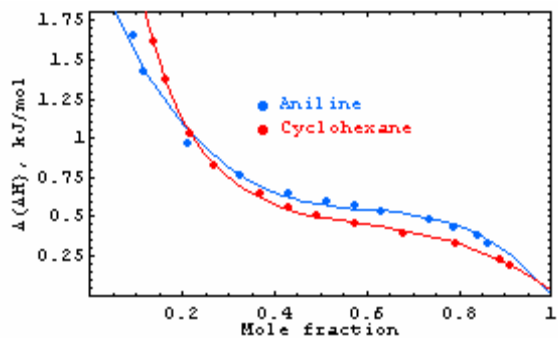

**Figure 7.** Variations of enthalpy of evaporation of components $\Delta(\Delta H)$ versus mole fractions in a mixture of Aniline and Cyclohexane at 343.15 K.

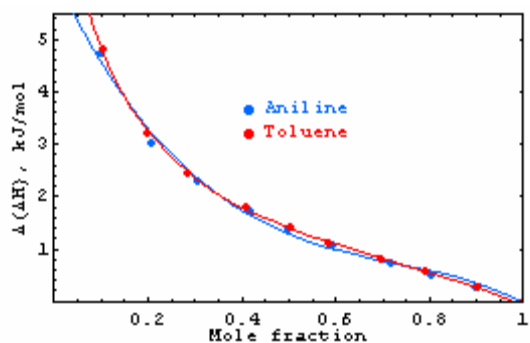

**Figure 8.** Variations of enthalpy of evaporation of components $\Delta(\Delta H)$ versus mole fractions in a mixture of Aniline and Toluene at 353.15 K.

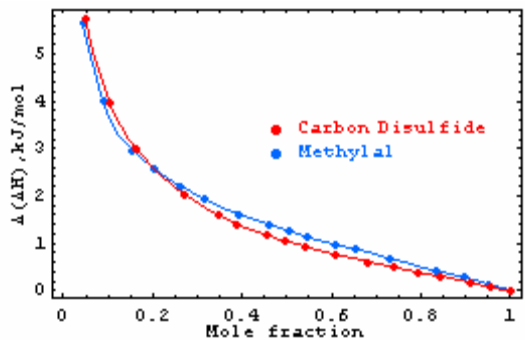

**Figure 9.** Variations of enthalpy of evaporation of components $\Delta(\Delta H)$ versus mole fractions in a mixture of Carbon Disulfide and Methylal at 308.32 K.

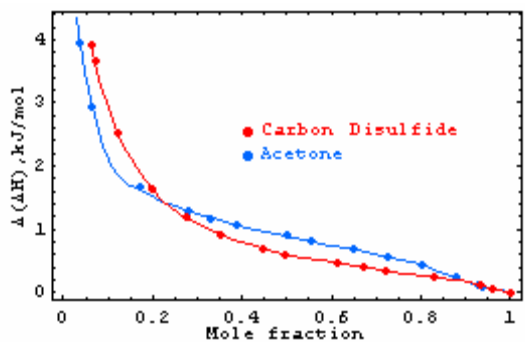

**Figure 10.** Variations of enthalpy of evaporation of components Δ(Δ*H*) versus mole fractions in a mixture of Carbon Disulfide and Acetone at 308.32 K.

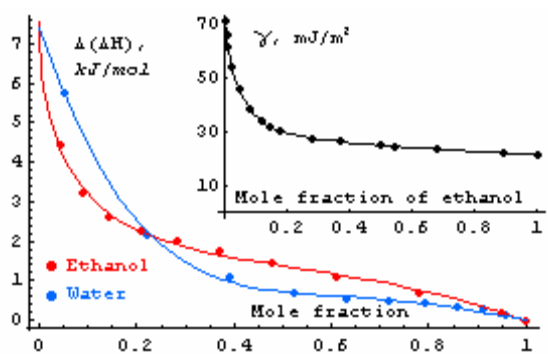

**Figure 11.** Variations of enthalpy of evaporation of components Δ(Δ*H*) versus mole fractions in a mixture of Ethanol and Water at 348.15 K. Insert. Surface tension of the ethanol-water mixture at 293 K.

Consider the variations of enthalpy in the limits of low and high concentrations.

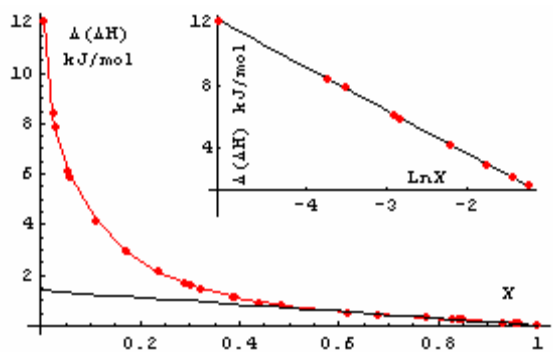

**Figure 12.** Variations of enthalpy of evaporation of Chloroform Δ(Δ*H*) versus its mole fractions *X* in a mixture of Chloroform and Ethanol at 308.15 K. Black line

shows that $\Delta(\Delta H)$ varies linearly at low concentrations of Ethanol. **Insert.** The initial points of $\Delta(\Delta H)$ (0<*X*<0.3) versus logarithm of mole fractions.

Figures 12 and 13 show that at high concentrations $\Delta(\Delta H)$ diminishes linearly with mole fractions, whereas in intervals of low concentrations (see Inserts to Figures 12 and 13) the values of $\Delta(\Delta H)$, as for the ideal solutions, decrease linearly with logarithm of mole fractions.

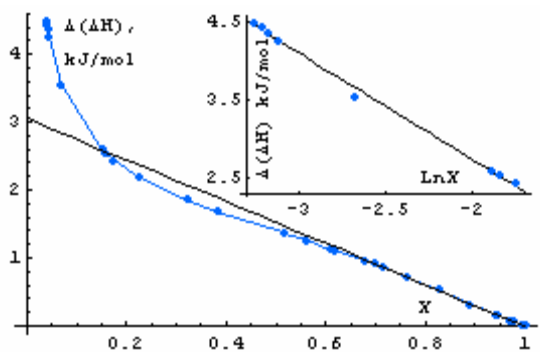

**Figure 13.** Variations of enthalpy of evaporation of Ethanol $\Delta(\Delta H)$ versus its mole fractions *X* in a mixture of Chloroform and Ethanol at 308.15 K. Black line shows that $\Delta(\Delta H)$ varies linearly at low concentrations of Chloroform. **Insert.** The initial points of $\Delta(\Delta H)$ (0<*X*<0.3) versus logarithm of mole fractions.

Since $\Delta(\Delta H)$ is supposed to be constant in a small temperature range, the values of $\Delta(\Delta H)$ obtained at the definite temperature can be used for extrapolating the equilibrium data to another temperature.

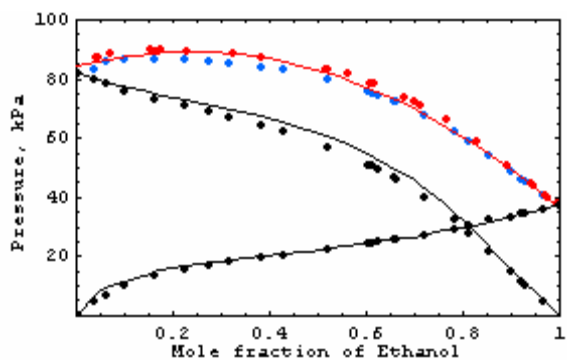

**Figure 14.** Partial and total vapor pressures of the Chloroform-Ethanol mixture at 328.15 K. Red points refer to the calculated total pressures; blue points indicate the experimental total pressures. Solid black lines designate the calculated partial pressures, while the black points correspond to their experimental values.

Figure 14 demonstrates the calculated and experimental values of total and partial pressures of the Chloroform-Ethanol mixture at $T$=328.15 K. The calculation is made proceeding from values of $\Delta(\Delta H)$ at 308.15 K (Figures 15-17). One may see that there is a reasonable agreement between the calculation and observations.

To clarify the reason of deviations of $\Delta(\Delta H)$ from the ideal solution behavior, one must take into account the distinction between thermodynamic properties of solutions, taken alone, and the enthalpy of evaporation. In the case of a bulk solution, the contribution of its surface layer to the average thermodynamic parameters is negligible, since the ratio of the number of molecules in the surface layer to the total number of molecules in the solution is close to zero; in the case of evaporation, *each of the bulk molecules* passes through the surface layer and, thus, its composition and properties affect the energy change.

**Model of evaporation**

Let us introduce a simple molecular model of evaporation, which takes into account the surface phenomena and can shed light on the effect of these latter on variations of enthalpy in solutions. This model has been earlier used to account for the effect of surface curvature on the enthalpy of condensation.[13-16] Since $\Delta(\Delta H)=\Delta H^s-\Delta H^0$ is calculated at the same temperatures of solutions and pure liquids, the kinetic energies of species in the solution, $E^s_{kin}$, and those in the pure liquid, $E^0_{kin}$, are equal to one another and only the changes of the potential energy, $\Delta E_{pot} \equiv E^0_{pot}-E^s_{pot}$, are significant.

Keeping in mind that for condensed phases the values $\Delta U$ and $\Delta H$ are practically identical, one obtains the sequence of equalities:

$$\Delta(\Delta H) \cong \Delta(\Delta U) = \Delta(\Delta E_{kin} + \Delta E_{pot}) = \Delta E_{pot}$$

(8)

As the distinction between condensation and evaporation is trivial, we shall retain the model described in earlier publications and continue the discussion in terms of internal energy of condensation.

Because $U$ is the state function, one can choose any convenient way between initial and final states for calculating $\Delta U$. In particular, imagine that condensation consists of three consecutive stages: in the first stage, each gas molecule is adsorbed on the surface and then, in the second and third stages, it penetrates the surface layer and, finally, into the interior of the liquid. The picture that emerges from the model is that of a quiescent liquid surface, while it is actually in the state of violent agitation on the molecular scale with individual molecules passing back and forth between the surface and the bulk regions on either side. As Adamson[17] writes, "under a microscope of suitable magnification, the surface region should appear as a fuzzy blur, with the average density varying in some continuous manner from that of the bulk phase to that of the vapor phase." It may appear that there is a conflict between our model and the reality. In this connection, the following should be taken into account. The model does not pretend to be the model of surface region. It just takes advantage of the path-independence of enthalpy and introduces the hypothetical intermediate states attributing to these latter the physical properties of real surfaces such as chemical compositions and surface tensions. Furthermore, it follows from eq 8, $\Delta(\Delta H) \approx \Delta E_{pot}$, that the changes of enthalpy may be described in terms of motionless molecules.

The first stage of the model is autoadsorption and its energy effect is the energy of autoadsorption. This term denotes adsorption of vapor on the surface of its own condensed phase (for example, water vapors on the surfaces of either ice or liquid water) when molecules only touch the surface without entering the surface layer. In the parlance of adsorption theory, the autoadsorption energy ε* is the energy of adsorption in the Henry limit (that is, at the limit of zero coverage); after the adsorption layer has been completed, each of adsorbed molecules interacts not only with species located beneath the layer but also with its lateral neighboring molecules. The energy of lateral interactions, ε*$_{lat}$, contributes to the total effect and determines a change of internal energy corresponding to the penetration in the surface layer. The energetic effect of the third stage when a molecule moves from the surface layer to the bulk liquid is equal in magnitude but opposite in sign to the total (excess) surface energy, $E_s$. The total surface energy is just defined as the energy gained by a molecule on being transferred from the bulk liquid to its surface;[7, 17-20] it is the excess energy of the molecules in the surface layer in respect to their energy in the volume. This value is in a close association with the free surface energy (or the surface tension), γ:

$$E_s = \gamma - T \frac{d\gamma}{dT}$$

(9)

where (-dγ/d$T$) is the excess surface entropy and $T$(-dγ/d$T$) is the quantity of latent heat absorbed in the reversible isothermal change of the surface area.[20] $E_s$ is nearly temperature independent in the wide range not too close to the critical temperature, $T_c$, but eventually drops to zero at $T_c$.[17] The interplay between the variation of internal energy on condensation, $\Delta U_{con}$, on one hand, and ε*, ε$_{lat}$, and $E_s$, on the other hand, is schematically depicted in Figure 15.

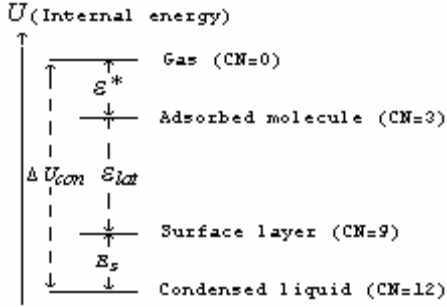

**Figure 15.** Schematic diagram of the energetic levels of molecules. $\Delta U_{con}$, $\varepsilon^*$, $\varepsilon_{lat}$, and $E_s$, are the energy of condensation, the autoadsorption energy, the energy of lateral interactions, and the excess surface energy, respectively. *CN* is the coordination number of a molecule in the case of a hypothetical closed packed liquid.

It is evident from Figure 15 that

$$\Delta U_{con} = -(\varepsilon^* + E_s + \varepsilon_{lat}^*) \tag{10}$$

By convention, $\varepsilon^*$ is positive, whereas $\Delta U$ on condensation is negative. As accepted in the theory of adsorption, the asterisk $^*$ indicates that the parameters corresponds to the well depth. Then, for evaporation one obtains

$$\Delta U_{ev} = \varepsilon^* + E_s + \varepsilon_{lat}^* \tag{11}$$

Applying equation 11 to the pure liquid and to the same liquid in a solution, one has

$$\Delta(\Delta H) \cong \Delta(\Delta U) = \Delta \varepsilon^* + \Delta E_s + \Delta \varepsilon_{lat}^* \tag{12}$$

Hence, the total variation of enthalpy of evaporation is the sum of the change of autoadsorption energy $\Delta \varepsilon^*$, the variation of energy of lateral interactions $\Delta \varepsilon_{lat}$, and that of the excess surface energy $\Delta E_s$. Thus, the energy of evaporation is thought of as the sum of energetic changes of three sequential stages: passage of molecules from the bulk liquid into the surface layer, exit of the molecules on the outer side of the interface, and the following desorption into the gas phase.

As an example, consider a model of a closed packed liquid. In the bulk and in the surface layer of such a liquid each of the molecules is surrounded by 12 and 9 nearest neighbors, respectively. A molecule adsorbed on the surface of close-packed spheres at the position of the minimum potential energy has only three nearest neighbors. One may see that the number of nearest neighbors lost by a molecule while moving from volume to surface (12-9) is equal to that on desorption into a gas phase (3-0). As the energetic properties are mainly determined by the interactions with the nearest neighbors, one might expect that the autoadsorption energy and the excess surface energy are identical:

$$E_s = \varepsilon^*  \qquad (13)$$

These argumentations were put forward by Stefan[7] and Skapski;[21] the general evidence for this relation is given in the Appendix.

Consider a numerical example. For any pure liquids one has

$$\Delta H_{ev} = 2E_s + \varepsilon_{lat}  \qquad (14)$$

In particular, for water at 298 K $\Delta H_{ev}$=44 kJ/mol, $E_s$=7.45 kJ/mol;[17] hence, $\varepsilon_{lat}$=29.1 kJ/mol; for ethanol $\Delta H_{ev}$=39.37 kJ/mol, $E_s$=5.61 kJ/mol[17] and $\varepsilon_{lat}$=28.15 kJ/mol. In the case of hypothetical closed packed liquid, $\Delta H$=4$E_s$ and $\varepsilon_{lat}$=2$E_s$. One may see that the main contribution to enthalpy of evaporation comes from the surface layer.

**Effect of adsorption in ideal and non-ideal solutions**

It should be particularly emphasized that eq 13 is valid for pure liquids when both the surface layer and the bulk liquid have the same chemical compositions; in the cases of the binary solutions, adsorption from the bulk to the interface changes the chemical composition of the surface layer and both the energy of autoadsorption and the total surface energy become different from those of the pure liquids. It is known that the components of solutions lowering their surface tensions are concentrated on the interfaces. For example, for the ethanol-water mixture the surface tension of the solution is lowered by the addition of ethanol (Figure 11. Insert) and varies at 293.15 K from that of water, 72.88 mJ/m$^2$, to $\gamma$ of pure ethanol, 22.39 mJ/m$^2$, the variations of $\gamma$ being practically completed at $X$>0.3. It is also known that molecules of ethanol are oriented on the solution surface so that the hydroxyl groups are turned toward the solution and the hydrocarbon groups are turned to the vapor phase.

Consider the ethanol-water mixture in the interval of concentration of ethanol from $\approx$0.3 to 1. Since $\gamma$ and, hence, the arrangement of the surface layer in this interval approaches that of ethanol, molecules of water in the gas phase "see in front of themselves" the ethanol surface. Hence, for these molecules of water (1) $\varepsilon^*$ is the energy of adsorption of water on the ethanol (hydrocarbon) surface and (2) $\varepsilon_{lat}^*$ is the interactions of water with the lateral ethanol environment. We recall that for the pure water these terms correspond to the water-water interactions. For ethanol molecules in the gas phase, both in the cases of the pure ethanol and the ethanol-water mixtures, $\varepsilon^*$ and $\varepsilon_{lat}^*$ in this interval are determined by the ethanol-ethanol interactions and, hence, in eq 12 $\Delta\varepsilon^*\approx0$ and $\Delta\varepsilon_{lat}^*\approx0$. From here, equation 12 in this interval predicts the different values of $\Delta(\Delta H)$ for evaporations of components: $\Delta(\Delta H)\approx\Delta E_s$ for molecules of ethanol and $\Delta(\Delta H) = \Delta\varepsilon^* + \Delta E_s + \Delta\varepsilon_{lat}^*$ for evaporation of water. One may see that adsorption in the surface layer disturbs the symmetry of interactions; variations of enthalpy cease to be dependent only upon the number of 1-2 interactions in the solution and are influenced also by the surface processes. Probably, it is one of the reasons that the plots of $\Delta(\Delta H)$ versus concentrations for water and ethanol do not coincide with one another resulting in the deviation from the ideal behavior. As a whole, such a phenomenon is expected to occur in the mixture with the different surface tensions of components.

Some of the supporting evidence for this thesis comes from Table 2 presenting the differences between the surface tensions of components of binary solutions at 293.15 K. It would be preferable to compare the properties of solutions with the difference of total surface energy $\Delta E_s$ expressed in units of J/mol, and not with $\Delta\gamma$ ($J/m^2$); but due to the scarce information on the surface area occupied by molecules in the surface layer, which is needed for calculating $\Delta E_s$ (J/mol), as the first approximation, we restrict ourselves to the comparison with $\Delta\gamma$ ($mJ/m^2$). In the cases of the ideal and ideal dilute solutions, $\Delta\gamma$ is less then 4.0; mixtures with $\Delta\gamma\approx5$ take the intermediate positions between ideal and non-ideal solutions: they may be treated, depending on the accuracy, as ideal dilute or as non-ideal solutions. At last, the liquid mixtures with $\Delta\gamma>8.8$ behave like the non-ideal solutions.

Table 2. Difference of surface tensions for the binary solutions

| solution | $\gamma$, $mJ/m^2$ | $\gamma$, $mJ/m^2$ | $\Delta\gamma$, $mJ/m^2$ | comment |
|---|---|---|---|---|
| Hexane-Heptane | 18.43 | 20.29 | 1.86 | Ideal or ideal diluted solutions |
| Aniline- Methyl aniline | 43.66 | 39.6 | 4.0 | |
| $CCl_4$-Benzene | 26.66 | 28.66 | 2.0 | |
| Ethyl acetate- $CCl_4$ | 23.9 | 26.66 | 2.76 | |
| Chloroform-Acetone | 28.86 | 23.7 | 5.16 | intermediate solutions |
| Ethyl bromide- Ethyl iodide | 24.16 | 29.5 | 5.34 | |
| Ethanol-Chloroform | 22.03 | 28.86 | 8.83 | Non-ideal solution |
| Aniline-Cyclohexane | 43.66 | 25.3 | 18.36 | |
| Aniline-Toluene | 43.66 | 28.53 | 15.13 | |
| Carbon Disulfide-Methylal | 31.38 | 21.4 | 10 | |

| Water-Ethanol | 72.75 | 22.03 | 50.52 | |
|---|---|---|---|---|

In the cases of ideal solutions with the similar values of $\gamma$ for both components, the effect of adsorption in the surface layer is negligible and the chemical compositions of the surface layer is close to that of the bulk liquid. Therefore, the surface phenomena in ideal solutions have no influence on the variations of $\Delta(\Delta H)$.

**Conclusions**

The equation derived on the basis of the fundamental Clapeyron equation establishes the relation between enthalpy of evaporation of components and their concentrations in the solution. The treatment of experimental observations shows that the variations of enthalpy are strongly influenced by non-ideality of solutions. It is supposed that the total change of enthalpy is equal to the sum of its variations in the bulk liquid and that on the interface, the latter determining, to great extent, the deviation from the ideal behavior.

Equation 5 had been earlier used to compute the equilibrium pressures over (i) the curved liquid surface[13-16] and over (ii) adsorbates (liquids) in the pores of adsorbents.[22, 23] The reference systems for these problems were the equilibriums over (i) the flat liquid surface and over (ii) the nonporous flat surface of adsorbent coved by adsorbate. Such an approach showed a very good correlation with experimental observations; the equilibrium pressures proved to be in the quantitative agreement with those predicted by density functional theory. In cases of curved surfaces and adsorbents, equation 5 was derived not only from the classical thermodynamics but also from statistical mechanics. The generality of approaches to resolving the apparently distinct problems, which is an inherent property of the thermodynamic relations, gives promise that equation 5 may be the basis for studying the concentration-dependent parameters also for other equilibrium systems.

**Appendix**

Consider the transformation of the surface layer. As the adsorption layer is being filled by adsorbate, it evolves into the new surface layer (Figure 16) where each of molecules interacts also with the lateral molecules.

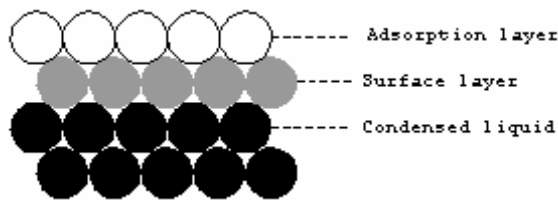

**Figure 16.** Schematic diagram of the process in the surface layer. Surface molecules (gray disks) covered by the adsorbed molecules (circles) experience the same environment as molecules within a bulk phase (black disks).

After the adsorption layer has been completed, the former surface molecules (gray disks) turns into the volume molecules: their coordination number (CN) increases to the CN of bulk molecules (black disks) and they become indistinguishable from these latter. Figure 16 illustrates a liquid with the surface layer of one molecular thick.

 In general, the thickness of the surface region is determined by the intermolecular forces, but it is independent of quantity of a liquid. Hence, to keep the surface layer thickness constant with the addition of the adsorption layer, the equal quantity of liquid from the surface region must be moved into the volume. This means that there is a one-to one correspondence between the phenomena and at constant $T$ the energy of autoadsorption is equal to $E_s$.